\documentstyle[twoside,fleqn,espcrc2,epsfig]{article}

\title{Photoproduction of $J/\psi$ and $\Upsilon$ in $pp$ and
$\overline{p} p$ Collisions}

\author{Spencer Klein\address[LBNL]{Lawrence Berkeley National
Laboratory, Berkeley, CA 94720, USA} and
Joakim Nystrand\address[Bergen]{Dept. of Physics, 
University of Bergen, N-5007, Bergen, Norway}
}
\begin{document}

\begin{abstract}

Exclusive vector meson photoproduction, $pp\rightarrow ppV$ and
$\overline pp\rightarrow\overline ppV$ occurs with significant rates
at hadron colliders.  The reaction can be used to study the gluon
distribution of protons.  Vector mesons may be produced with either
proton as a target; because of interference between the two
production channels, the $p_T$ spectra of vector mesons produced in
$pp$ and $\overline pp$ collisions are quite different.  Because of
the unique event signature, vector meson photoproduction can be
separated from hadroproduction events, despite the small ratio of
cross sections.  We consider production of $J/\psi$ and $\Upsilon$ at
RHIC, the Tevatron and the LHC.

\vspace{1pc}
\end{abstract}

\maketitle
\section{Introduction}

Most recent studies of photoproduction have been done with fixed
target, real photon beams, or in $ep$ collisions at the HERA collider,
where the electromagnetic field of the electron is a source
of photons.  However, collisions at HERA are limited to photon-proton
center of mass energies of about 200 GeV, and no higher energy
photoproduction facilities are planned.

However, photoproduction can also be studied at $pp$ and $\overline p
p$ colliders.  These collisions offer some advantages over $ep$
collisions.  The Fermilab Tevatron can reach somewhat higher $\gamma
p$ energies than HERA, and the LHC can reach more than an order of
magnitude higher in energy than HERA.  Although photoproduction is a
smaller fraction of the total cross section at $pp$ colliders than at
HERA, higher machine luminosities can at least partly make up for
this.  Finally, the CP antisymmetric $pp$ and CP symmetric $\overline
p p$ initial states allow for some interesting interference effects.

Similar reactions are already studied at heavy ion colliders
\cite{reviews}.  The STAR collaboration at RHIC has measured the cross
section for $Au + Au \rightarrow Au + Au + \rho^0$ \cite{STARrho}, and
a significant photonuclear and two-photon physics program is planned
for the LHC \cite{felix,cms}.  Cross sections for $J/\psi$
\cite{usPRC,strikman} and $\Upsilon$ \cite{strikman2} have been
calculated.  Photonuclear interactions with heavy ions are limited to
lower energies than for proton collisions.

Here, we consider photoproduction of $J/\psi$ and $\Upsilon$ at RHIC,
the Tevatron, and the LHC.  Although it runs at a somewhat lower
energy than the Tevatron, RHIC collides polarized protons, allowing
for photoproduction studies with polarized targets.

After introducing the ideas behind photoproduction, we will show the
rates, rapidity distributions and $p_T$ spectra for these cases.  We
will also note how this reaction can be used to measure gluon parton
distributions down to $10^{-3}$ at the Tevatron and $2\times 10^{-4}$
at the LHC.  After discussing the production rates, we will briefly
discuss the experimental feasibility of the measurement.  Although
we focus on vector meson production, other photoproduction reactions
should also be accessible at proton colliders.

$J/\psi$ photoproduction in $pp$ collisions has been considered
elsewhere \cite{khoze}.  That calculation used a very different
approach, focusing on the energy lost by the protons.  This makes
analytic or numerical comparisons between their result and ours very
difficult.

\section{Photoproduction}

Photoproduction rates depend on two factors: the photon flux, and the
photon-proton cross section.  The cross section is
\begin{equation}
\sigma(pp\rightarrow ppV) = 2 \int dk {dn \over dk} \sigma(\gamma p
\rightarrow Vp).  
\end{equation}
where $k$ is the photon energy, $dn/dk$ is the photon flux, and
$\sigma(\gamma p\rightarrow Vp)$ is the corresponding photoproduction
cross section.  The '2' accounts for the fact that either
proton/anti-proton can be the emitter or the target.

The photon flux is given by the Weizs\"acker-Williams method of
virtual photons.  For high energy photons, there is a need to consider
the proton internal structure.  We use the photon spectrum given by
Drees and Zeppenfield \cite{DZ}
\begin{equation}
{dn\over dk}\! =\! {\alpha\over 2\pi k}
\big[1 + (1-{2k\over\sqrt{s}})^2\big]\!
\big(\ln{A} - {11\over 6} + {3\over A} -{3\over{2 A^2}} +{1\over 3A^3}\big)
\end{equation}
where $\alpha$ is the fine structure constant and
\begin{equation}
A = 1 + {0.71{\rm GeV}^2\over Q^2_{min}} .
\end{equation}
Equation 2 is derived using a proton form factor with 
$Q^2_{min} = k^2/(\gamma^2 (1 - 2k/\sqrt(s)))$,
where $\gamma$ is the Lorentz boost of the beam and $\sqrt{s}$ is the
proton-proton center of mass energy.  This spectrum is close to that
of a point charge with a minimum 0.7 fm impact parameter.

For exclusive vector meson production, the photoproduction reaction
must not be accompanied by hadronic interactions.  To explore this, we
consider an alternative photon flux.  We eliminate the proton form
factor, treating the proton as a point charge, and instead require
$b_{min}=1.0$ fm.  This reduces the vector meson production by about
20\%; half the difference between these calculations is a reasonable
estimate of the systematic uncertainty.  A more detailed calculation,
including the hadronic interaction probability as a function of $b$,
should lead to a more accurate photon flux, with smaller errors.
The photon flux in $pp$ and $\overline{p}p$ collisions can also be
studied by considering other photoproduction and two-photon
interactions.

The photoproduction cross sections, $\gamma p\rightarrow Vp$ depend on
the gluon density in the proton\cite{ryskin}.  At mid-rapidity,
$J/\psi$ production is sensitive to gluon with $x$ down to
$1.5\times10^{-3}$ at the Tevatrony and $2\times10^{-4}$ at the LHC.
Away from mid-rapidity, significantly lower gluon energies can be
probed.

For cross section estimates, we use data from HERA. Of course, this
requires some extrapolation at the Tevatron, and significant
extrapolation at the LHC, but that's why the measurement is
interesting.

We parameterize the $J/\psi$ cross section\cite{HERApsi,H1}
\begin{equation}
\sigma(\gamma p\rightarrow J/\psi p) = 1.5 W^{0.8} (\rm GeV) {\rm nb}
\end{equation}
where W is the $\gamma p$ center of mass energy.  Up to $W\approx 200$
GeV, this is well measured.

The $\Upsilon(1S)$ cross section is parameterized\cite{H1,HERAups}
\begin{equation}
\sigma(\gamma p\rightarrow \Upsilon(1S) p) = 0.06 W^{1.7} (\rm GeV) {\rm pb}.
\end{equation}
Neither the absolute values of the cross section nor the energy
dependence are well known, so there is considerable room for an
improved measurement.  The ratio of $\Upsilon(1S)$ to $\Upsilon(2S)$
to $\Upsilon(3S)$ is also unknown; the ZEUS and H1 collaborations
assumed that about 70\% of the signal was from $\Upsilon (1S)$.  This
ratio seems to be measurable at the Tevatron.

The photon energy can be determined from the rapidity of the produced
vector meson, $y$:
\begin{equation}
y = \ln{({2k\over M_V})}
\end{equation}
One complication is that either beam particle is equally likely to
produce the photon; the cross sections for these two possibilities are
added.  Unambiguous determination of the photon energy is possible
only at $y=0$; more sophisticated analyses are needed to determine the
cross section for other photon energies.  Two promising approaches to
moving away from $y=0$ are to use lower energy data to determine the
cross section for the lower energy photon, and to measure the ratio
from the two directions by measuring the interference fraction, as
discussed below.  It may also be possible to use Roman pots or other
forward taggers to track the scattered protons; the proton with the
larger $p_T$ is very probably the target. This approach could at least
partially resolve the two-fold direction ambiguity, and simplify gluon
distribution measurements away from $y=0$.

\begin{figure}[t]
\setlength{\epsfxsize=2.9 in} 
\setlength{\epsfysize=2.1 in}
\centerline{\epsffile{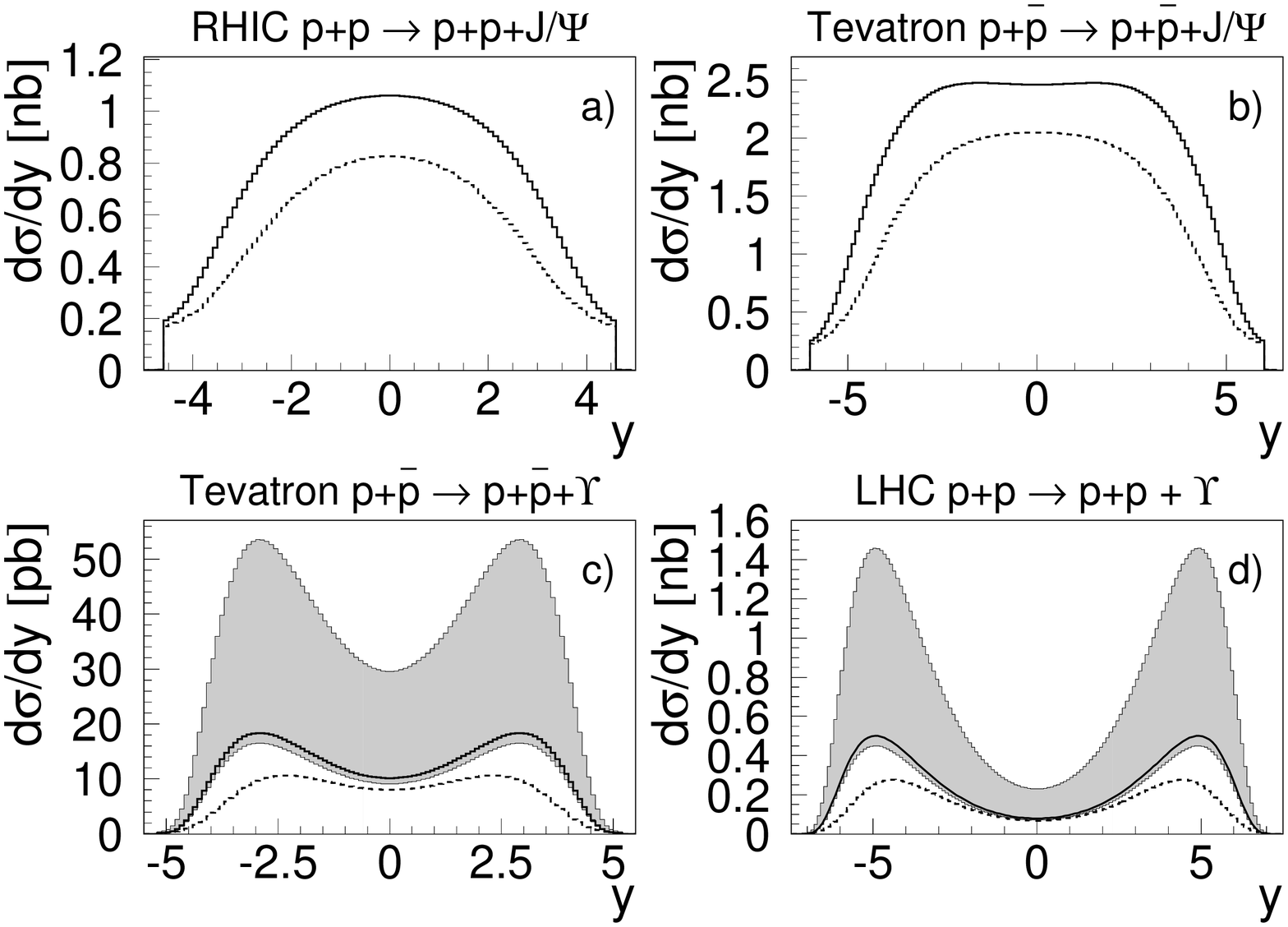}
}
\caption[]{Rapidity distributions, $d\sigma/dy$ for (a) $J/\psi$
production in 500 GeV $pp$ collisions at RHIC, (b) $J/\psi$ production
in 1.96 TeV $\overline pp$ collisions at the Tevatron, (c) $\Upsilon(1S)$
production in 1.96 GeV $\overline pp$ collisions at the Tevatron, and
(d) $\Upsilon(1S)$ production in 14 TeV $pp$ collisions at the LHC.  The
solid histograms are as described in the text.  The dashed histogram
is done with $b_{min} = 1$ fm, but without the proton form factor
(i.e. for a point charge).  The shaded regions in (c) and (d) show the
cross section uncertainty from the HERA measurements.  }
\end{figure}

Here, we consider 3 facilities: $pp$ collisions at $\sqrt{s}=500$ GeV
at RHIC, $\overline p p$ collisions at 1.96 TeV at the Tevatron (the
results shown at the Small-x and Diffraction 2003 were at
$\sqrt{s}=1.8$ TeV), and $pp$ collisions at 14 TeV at the LHC.  We
assume luminosities of $10^{31}$/cm$^2$/s, $2\times 10^{32}$/cm$^2$/s
and $10^{34}$/cm$^2$/s respectively.  With these assumptions, we find
the $d\sigma/dy$ shown in Fig. 1.  Table 1 summarizes the total cross 
sections and rates for $J/\psi$ and $\Upsilon$ production.

Lighter vector mesons are also produced prolifically at $pp$ and
$\overline pp$ colliders.  The cross sections for $\rho^0$, $\omega$
and $\phi$ production should be several orders of magnitude larger
than for the $J/\psi$.

\begin{table}[t]
\begin{tabular}{l|rr|rr}
Machine& \multicolumn{2}{c|}{$J/\psi$}  & \multicolumn{2}{c}{$\Upsilon$}\\
& \multicolumn{1}{c}{$\sigma$} 
& \multicolumn{1}{c|}{Rate} 
& \multicolumn{1}{c}{$\sigma$} 
& \multicolumn{1}{c}{Rate} \\
\hline
RHIC     & 7.0 nb & $7.0\!\times\!10^{4}$    & 12 pb  & 120 \\
Tevatron& 23  nb & $4.6\!\times\!10^{7}$    & 120 pb & $2.4\!\times\!10^5$ \\
LHC      & 120 nb & $1.2\!\times\!10^{10}$ & 3.5 nb & $3.5\!\times\!10^8$ \\
\hline
\end{tabular}
\caption[]{Total cross sections and rates for production of the $J/\psi$ and
the $\Upsilon$.  The rates are for $10^7$ s of running at the Tevatron
and the LHC.  For RHIC, where most of the running time is devoted to
heavy ions, $10^6$ s is used.}
\end{table}

\section{Interferometry and $p_T$ spectrum}

The vector meson $p_T$ depends on the $p_T$ of the photon and the
$p_T$ acquired when the vector meson is produced; the latter is
dominant in $pp$ collisions.  The $p_T$ of the photon depends on the
form factor of the proton/anti-proton:
\begin{equation}
{d^3N_\gamma\over d^2k_T dk} =
{\alpha F^2(k_T^2 + k^2/\gamma^2) k_T^2
\over
\pi k (k_T^2 + k^2/\gamma^2)^2}
\end{equation}
where the form factor 
\begin{equation}
F(Q^2) = { 1\over(1+Q^2/0.71 {\rm GeV}^2)^2}
\end{equation}
is the same one used previously.  Here, $k_T$ is the transverse
momentum of the photon. The $p_T$ from the vector meson production is
just given by the form factor.  We assume that the photon $p_T$ and
that obtained in the scattering are randomly oriented.

The final ingredient in the $p_T$ spectrum is an interference term
\cite{interfere}.  The two possibilities, proton \#1 emitting a photon
which scatters off proton \#2, and vice versa, are indistinguishable,
and so the amplitudes add.  The cross section for a given impact parameter, 
$b$, is
\begin{equation}
\sigma = |A_1 \pm A_2\exp{(ip_T\cdot b)}|^2
\end{equation}
where $A_1$ and $A_2$ are the amplitudes for production for the two
possibilities.  The exponential is a propagator to account for the
phase difference between the two positions.  At $y=0$, $A_1=A_2$ and
the cross section simplifies to
\begin{equation}
\sigma = \sigma_0 [(1 \pm \cos{(p_T\cdot b)}]
\end{equation}
The sign of the interference term depends on the symmetry of the
system.  For $pp$ (and $AA$) colliders, one possibility can be
transformed into the other via a parity transformation.  Vector mesons
are $J^{PC} = 1^{--}$, {\it i.e.} negative parity, so the minus sign
applies.  For $\overline p p$ colliders, the requisite transformation
is a CP transform, so the plus sign applies.  

Of course, the impact parameter $b$ is unmeasurable, so the
$p_T$ spectrum is obtained by integrating over $b$.  This
washes out the interference except for $p_T < \hbar/\langle b\rangle$
where $\langle b\rangle$ is the mean impact parameter.  However,
for $pp$ collisions, the mean $p_T$ is not too different from
$\hbar/\langle b\rangle$, so the interference has a large
effect on the $p_T$ spectrum.

Figure 2 shows the $p_T$ distributions for $\Upsilon$ production at
RHIC and the LHC.  Because of the different sign of the interference,
they look quite different.
\begin{figure}
\centerline{
\setlength{\epsfxsize=2.9 in} 
\setlength{\epsfysize=2.1 in}
\epsffile{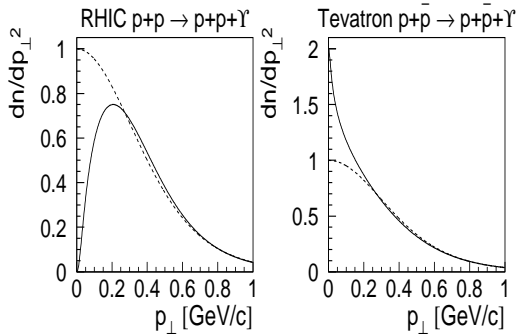}}
\caption[]{The $p_T$ spectra at mid-rapidity for photoproduced
$\Upsilon$ for (left) RHIC and (right) the Tevatron.  The solid curves
are the predicted spectrum, while the dashed curves are the spectra
without interference.  The latter curves would be more applicable at
large rapidity.}
\end{figure}

\section{Experimental Feasibility}

As Table 1 shows, the $J/\psi$ rates are high at all 3 accelerators.
Even after accounting for leptonic branching ratios and reasonable
acceptance estimates, the signals should be large.  The $\Upsilon$
rates are high at the Tevatron and the LHC, but, because of the small
leptonic branching ratios, are marginal at RHIC.

At RHIC and the Tevatron, the $J/\psi$ and $\Upsilon$ photoproduction
cross sections are both about 0.1\% of the corresponding
hadroproduction cross sections \cite{PHENIX,CDF}.  However, because
photoproduction is exclusive, $p + p \rightarrow p + p + V$, it should
not be difficult to separate the signal from hadronic backgrounds.
Three characteristics appear useful in separating the signal: the
meson $p_T$, the presence of rapidity gaps, and the detection of the
scattered protons.  Here, we focus on the $\Upsilon$.

Hadronically produced vector mesons typically have $p_T\approx
M_V$.  A cut on $p_T < 1$ GeV/c eliminates 94\% of the hadronic
background, while leaving almost all of the photoproduction.

Isolation cuts based on rapidity gaps can provide the remaining
separation power.  In a hadronic interaction, the probability of
having a rapidity gap with with $\Delta\eta$ is $\exp(-\Delta\eta\cdot
<dn_{ch}/dy>)$ where $<dn_{ch}/dy>$ is the mean charged particle
multiplicity.  At RHIC, $<dn_{ch}/dy> = 3$ at $y=0$, rising to 3.9 at
the Tevatron.  So, a single gap with width $\Delta\eta = 3$ will
reduce the background by a factor of $10^{-4}$ at RHIC and $10^{-5}$
at the Tevatron, {\it i.e.}  well below the signal level.  This is
well within the capability of large detectors.  In fact, the situation
is better than this, since there will be gaps on both sides of the
vector meson.

Exclusive $J/\psi$ production has been seen by the CDF Collaboration 
in $\overline pp$ collisions at the Tevatron\cite{Angela}. We believe 
that the origin of these exclusive events is photoproduction in peripheral 
collisions, as described above. 

If they were needed, Roman pots could be used to detect the scattered
protons, gaining cleanliness at some cost in detection efficiency.
One attractive feature of using Roman pots is that if the proton $p_T$
can be measured, it should provide some information on which proton
emitted the photon, and which was the scattering target.  

Cuts on rapidity gaps are less effective in rejecting other
diffractive interactions.  However, vector mesons have the wrong
quantum numbers to be produced in double-Pomeron interactions.  One
background is double-Pomeron production of the $\chi_{c0}$ or
$\chi_{b0}$, decaying to $\gamma J/\psi$ or $\Upsilon\gamma$, where
the photon is missed.  At the Tevatron, the double-Pomeron cross
sections for the $\chi_{c0}$ and $\chi_{b0}$ are about 600-735 nb and
0.1-0.9 nb respectively\cite{khoze2,yuan}.  After accounting for the
branching ratio, these backgrounds are small for the $J/\psi$, and
very probably small for the $\Upsilon$.  Also, the $p_T$ for
double-Pomeron produced mesons is considerably larger than for
photoproduction.

The photoproduction cross section rises with beam energy much more
rapidly than the soft Pomeron trajectory, so the ratio of
photoproduction to hadronic interactions, both diffractive and
non-diffractive, should grow as the beam energy increases.  This is
why photoproduction could not be observed in lower energy, fixed
target, experiments.  Conversely, it should be even more important at
the LHC.

\section{Conclusions}

The rates for photoproduction of $J/\psi$ and $\Upsilon$ mesons in
$pp$ and $p\overline p$ collisions at RHIC, the Tevatron and the LHC
are high, and these reactions can be used to measure the gluon content
of the proton at low $x$ values.  The $p_T$ spectrum of the produced
vector mesons are quite different at $pp$ and $p\overline p$
colliders.  The hadronic backgrounds to these reactions can be
controlled by selecting events with low $p_T$ vector mesons
in an otherwise empty detector.  

Photoproduction of other channels may also be of interest at hadron
colliders.  The cross sections for different photoproduction channels
rise rapidly with beam energy, in contrast to most diffractive channels.
So, as accelerator energy progresses, photoproduction will become more
and more important. 

It is a pleasure to acknowledge useful conversations with Angela Wyatt
and Mike Albrow.  We thank the Mike Albrow, Christophe Royon and
Maxine Hronek for organizing an enjoyable conference.  This work was
supported by the U.S. DOE under contract number DE-AC-03-76SF00098.

\end{document}